\documentclass[manuscript,showpacs,preprintnumbers,amsmath,amssymb,prb]{revtex4}
\usepackage{graphicx}
\usepackage{dcolumn}
\usepackage{bm}

\begin{document}

\title{Modeling semiconductor thermal properties. The dispersion role.}

\author{Damian Terris}
 \email{damian.terris@let.ensma.fr}
\affiliation{
Laboratoire d'Etudes Thermiques CNRS UMR 6608, Ecole Nationale Sup\'erieure de M\'ecanique et d'A\'erotechnique, University of Poitiers\\
ENSMA B.P 40109\\
86961 Futuroscope Cedex, France
}

\author{Karl Joulain}
 \email{karl.joulain@let.ensma.fr}
\affiliation{
Laboratoire d'Etudes Thermiques CNRS UMR 6608, Ecole Nationale Sup\'erieure de M\'ecanique et d'A\'erotechnique, University of Poitiers\\
ENSMA B.P 40109\\
86961 Futuroscope Cedex, France
}

\author{David Lacroix}
 \email{David.Lacroix@lemta.uhp-nancy.fr}
\affiliation{
Laboratoire d'Energetique et de Mecanique Theorique et Appliquee\\
Nancy Universite\\
54506 Vandoeuvre Cedex, France
}

\author{Denis Lemonnier}
 \email{denis.lemonnier@let.ensma.fr}
\affiliation{
Laboratoire d'Etudes Thermiques CNRS UMR 6608, Ecole Nationale Sup\'erieure de M\'ecanique et d'A\'erotechnique, University of Poitiers\\
ENSMA B.P 40109\\
86961 Futuroscope Cedex, France
}

\date{\today}

\begin{abstract}
We study heat transport in semiconductor nanostructures by solving the Boltzmann Transport Equation (BTE) by means of the Discrete Ordinate Method (DOM).
Relaxation time and phase and group velocitiy spectral dependencies  are taken into account.
The Holland model of phonon relaxation time is revisited and recalculated from dispersion relations (taken in litterature) in order 
to match bulk silicon and germanium values.
This improved model is then used to predict silicon nanowire and nanofilm thermal properties in both ballistic and mesoscopic regimes.

\end{abstract}
\maketitle

\section{\label{sec:level1}Introduction}

Semiconductors are often used in high technology due to their mechanical (crystal structure) and electrical (band gaps) properties. 
They are found in many domains such as nano-junctions \cite{Cui01}, transistors \cite{Wu04}, or in solar panels \cite{Springer04}, 
and are even used in medicine \cite{Kovalev05}. 
The increasing use of semiconductor micro/nano-structures has brought, nowadays, a good knowledge on charge transport in spite of heat transfer 
effects at very short scales\cite{Kim07}. 
The scope of this paper is to contribute to the charasterisation of semiconductor nanowire and nanofilm thermal properties.   
Semiconductor nanowires can be found for example in transistors \cite{ChenJ03} and semiconductor nanofilms in solar cells.

At these low scale, Fourier's law may give an inaccurate description of heat transfer through these nanostructures. 
One has to deal with more fundamental physics\cite{Cahill03}.
Several approaches as lattice dynamic simulation \cite{Gomes06,Chantrenne05,Mingo03,Ju99} or solving the 
Boltzmann Transport Equation\cite{Yang05nl,Chen01,Volz01,Goodson96,Lacroix05,Mazumder01} (BTE) 
have tried to explain experimental measurements \cite{Asheghi02,Li03}. 
In this paper, the BTE, which is a statistical physics equation, 
is solved by means of the discrete ordinate method (DOM) which is widely used in thermal radiation \cite{Modest03,Lemonnier07}.
We employ an improved model for describing phonon conduction by taking into account spectral dependancy, 
found in velocities and relaxation times, using dispersion curves.

At macroscopic scale, conductive heat transfer is governed by the heat equation:
\begin{equation}\label{diffusion}
\frac{\partial T(\mathbf{r},t)}{\partial t}=\nabla\left[k\nabla T(\mathbf{r},t)\right],
\end{equation}
where $k$ is the thermal conductivity. 
This equation can be derived in the framework of statistical physics considering that individual heat carriers (here phonons) 
are submitted to a random walk \cite{Einstein05}. 
In this diffusive regime, where collisions are numerous, the heat flux is related to the temperature gradient by Fourier's law :
\begin{equation}\label{Fourier}
\varphi(\mathbf{r},t)=-k\nabla T(\mathbf{r},t).
\end{equation}
When heat carriers do not undergo collisions, they cross the system like photons in thin optical layer. 
Heat carriers are then said to have a ballistic flight. A radiative-like form describes the steady state solution :
\begin{equation}\label{ballistic}
\varphi=\sigma (T^4_1-T^4_2).
\end{equation}
When phonon mean free path is of the same order as the system length, then some of them undergo collisions, 
while others cross the structure without any interaction. The system is in the mesoscopic regime.

Theoretically, as long as wave effects are negligible, the BTE applies to the ballistic, mesoscopic and diffusive regimes.
 A key feature of this approach is to introduce a specific intensity for phonon\cite{Kittel04}, 
similar to photon intensity used in radiative transfer \cite{Majumdar93}. 
Thus, the energy flux per apparent surface unit at point $r$ in direction $\mathbf{u}$ reads:
\begin{equation}
I_{\omega,p}(\mathbf{r},\mathbf{u})=\hbar\omega n_{\omega}(\mathbf{r},\mathbf{u}) \frac {V_g}{4 \pi},
\end{equation}
with $\hbar$ the Planck constant, $n_{\omega}(r,\mathbf{u})$ the phonon density at position $r$ going in direction $\mathbf{u}$ and $V_g$ 
the heat carrier group velocity. 
The BTE collision term is classicaly approached by the single relaxation time approximation and, therefore, in steady state, 
the transport resolution reads:
\begin{equation}\label{Boltzmann}
 \mathbf{u}\cdot \nabla I_{\omega,p}=\frac{I^0_{\omega,p}-I_{\omega,p}}{V_g \tau_{\omega,p}},
\end{equation}
where $I^0_{\omega,p}$ is the intensity at local equilibrium and $\tau_{\omega,p}$ the relaxation time.

\section{Numerical resolution}
\subsection{Nanostructure geometry modelling}\label{sec_geometry}

\begin{figure} [h]
\begin{center}
\includegraphics[width=8.6cm]{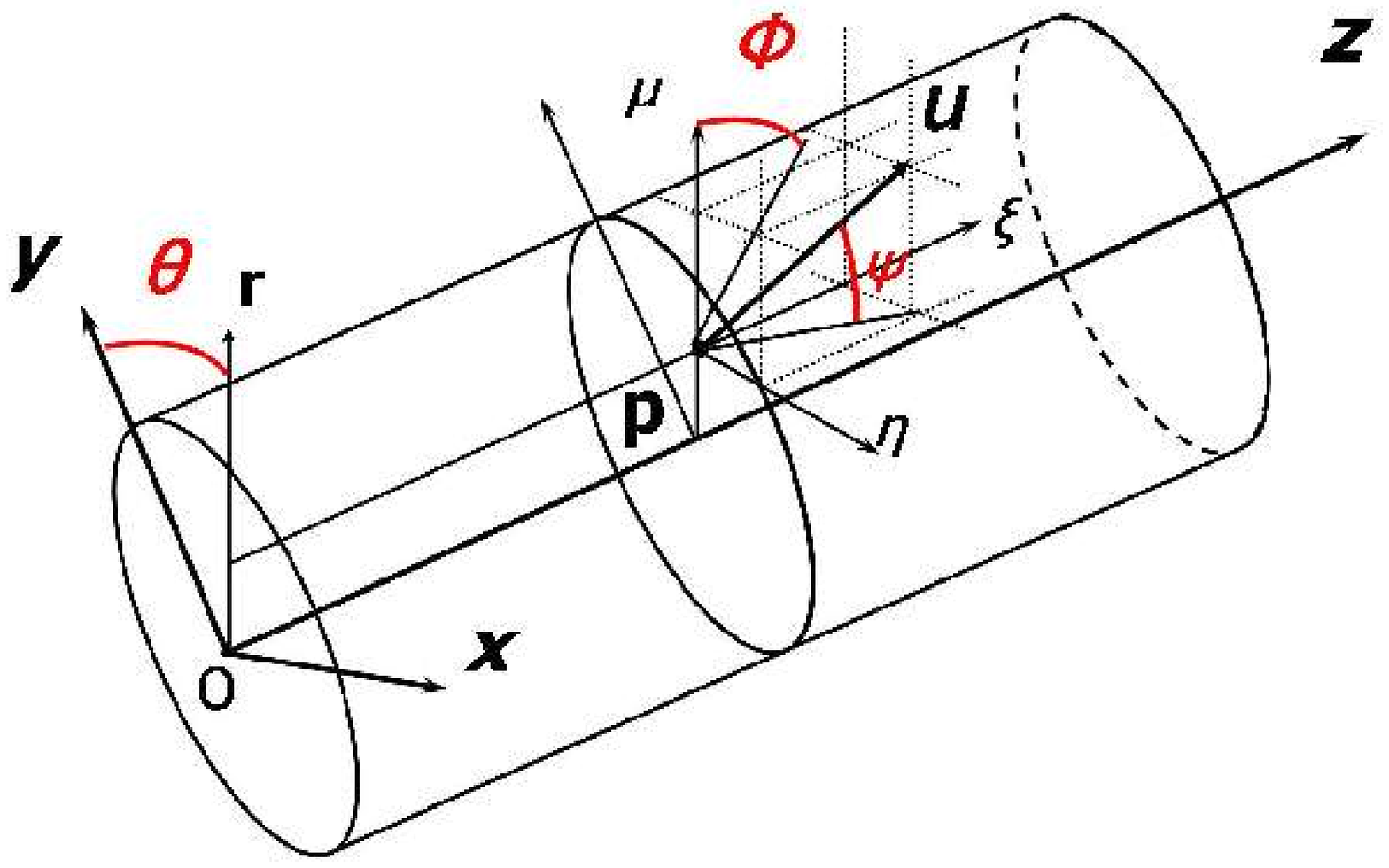}
\end{center}
\caption{(Color online) Geometry representation}
\label{geometry}
\end{figure}

We here consider nanofilms and nanowires made of semiconductor material.
The wire geometry refers to a cylinder whose length ($L$) is much larger than its diameter ($D$), $L>>D$. 
Cylindrical coordinates are obviously well adapted to this case. 
Films may also be viewed as cylinders in the limiting case where their thickness $L$ is much lower than the diameter $D$. 
Therefore, BTE in cylindrical coordinates still applies, it reads, for an axisymmetric problem\cite{Fiveland82}:
\begin{equation}\label{Boltz_cyl}
 \frac{\mu}{r} \frac{\partial {(r I_{\omega,p} )} }{\partial {r} }+\xi \frac{\partial {I_{\omega,p}} }{\partial {z} }
-\frac{1}{r} \frac{\partial {(\eta I_{\omega,p} )} }{\partial {\phi} }+\kappa_{\omega,p} I_{\omega,p}=\kappa_{\omega,p}I^0_{\omega}(T),
\end{equation}
where $\kappa_{\omega,p} = \frac {1}{V_g \tau_{\omega,p}}$ is the extinction coefficient and where $\mu$, $\eta$ and $\xi$ (Eqn.\ref{cosine}) 
are the direction cosines of the propagation direction $\mathbf{u}$ (fig. \ref{geometry}). 
They are linked to polar angles $\phi$ and $\psi$ characterising $\mathbf{u}$ in a local frame by:
\begin{equation}\label{cosine}
\left\{
\begin{array}{l}
\mu=\cos(\phi)\sin(\psi) ,\\
\eta=\sin(\phi)\sin(\psi) ,\\
\xi=\cos(\psi).
\end {array}
\right.
\end{equation}
The term $\frac {\partial}{\partial \phi}$ represents the angular redistribution and the variation over $\theta$ (fig.\ref{geometry}) is not 
written since $\frac {\partial}{\partial \theta}=0$ in an axisymmetric problem.

Figure \ref{geometry} shows a typical geometry under consideration. 
The two ending sections are always considered as black surfaces set at different temperatures:
 $T_{hot}$ on the left side and $T_{cold}$ on the right side ($T_{hot}>T_{cold}$). 
They emit phonons with a blackbody intensity at their wall temperature (Eqn.\ref{black}).
\begin{equation}
\label{black}
I^0_{\omega,p}(T_{Wall})=\frac{\hbar\omega^3}{8\pi^3V_p^2[\exp(\hbar\omega/k_bT_{Wall})-1]},
\end{equation}
where $V_p$ is the phase velocity and $T_{Wall}$ is either $T_{hot}$ or $T_{cold}$ .

The lateral (cylindrical) surface is treated as adiabatic, which amounts to consider pure reflection at the wall.
In the wire configuration, wall reflection is in general partly specular and partly diffusive.
In the film case, it is always specular so that there is no lateral boundary influence.
Condition on the cylinder axis is due to symmetry specular reflection.

\subsection{Discrete Ordinate Method - DOM}

\begin {figure}[h]
\begin{center}
\includegraphics[width=8.6cm]{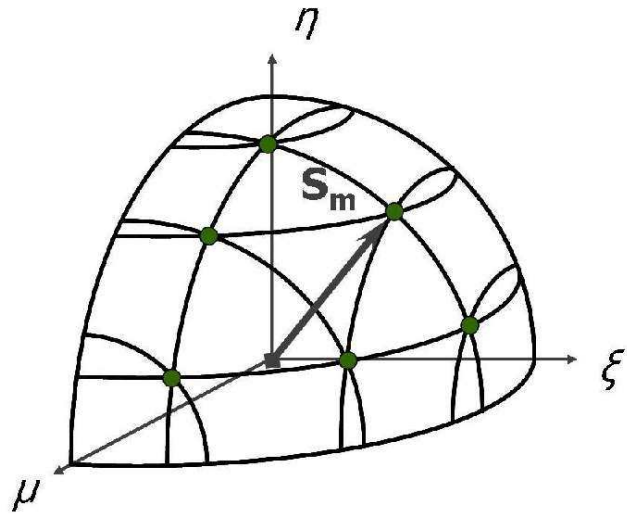}
\end {center}
\caption{(Color online) $S_6$ quadrature (first octant)}
\label{quad}
\end{figure}

The DOM is based on the selection of a finite set of propagation directions ($\mathbf{s}_m$, $m=1,\ldots,M$) and corresponding weights 
($w_m$)\cite{Jendoubi93,Balsara01,Koch04}. $S_N$ quadrature (fig. \ref{quad}) is the most commonly used. 
It is constructed with a maximum of symmetry requirements (to avoid directional bias) and the weight selection rules tend to preserve the exact values 
of some direction cosine key moments.

Boundary condition for any discrete direction leaving adiabatic surfaces are expressed as:
\begin{eqnarray} \label{reflection}
I_m(x_P)& = & \frac{\rho}{\pi} \sum_{m' \;{\rm if }\; \mathbf{s}_m'\cdot\mathbf{n} < 0 } w_{m'}I_{m'}(x_P)|\mathbf{s}_m'\cdot\mathbf{n}|  \nonumber\\
	    & + &(1-\rho)I_{\hat{m}}(x_P),
\end{eqnarray}
where, due to $S_N$ set symmetries, if $\mathbf{s}_m$ belongs to the quadrature, so does the specularly reflected direction $\mathbf{s}_{\hat{m}}$.
$\rho$ is the ratio of diffuse to specular reflection ($\rho=0$ yields pure specular reflection).

In this equation, as in the rest of the text, spectral indices $\omega$ and $p$ are omitted for sake of clarity.

In this work, $S_8$ quadrature is employed.
It involves $80$ directions, but thank to the symmetry only $40$ directions are used, with their associate weight doubled.

\subsection{Numerical procedures}

\begin {figure}[h]
\begin{center}
\includegraphics[width=8.6cm]{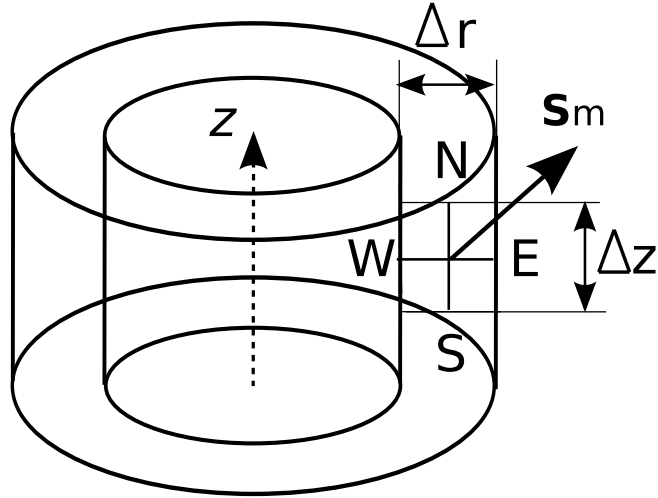}
\end {center}
\caption{ Four principle cardinal directions surrounding point $P$ with a given propagation direction $\mathbf{s}_m$. 
          $W$ and $S$ are the known values. Their positions are fixed in opposition to $\mathbf{s}_m$ propagation direction.}
\label{cardinal}
\end{figure}

The DOM is widely used to model radiative heat transfer and has already been employed to resolve steady or unsteady BTE 
for phonons\cite{Lemonnier07,Fiveland82}. 
Following Lathrop's guidelines \cite{Lathrop69}, 
a variable weight scheme (Eqn. \ref{diamond}) is used to relate cell-face intensities to the central value $I_{m,P}$ in a node $P$, 
for a given propagation direction $m$, at a given frequency $\omega$ and polarization $p$. 
Considering phonon propagation directions $\mathbf{s}_m$ represented in figure \ref{cardinal}, the discretized intensities will be written as:
\begin{eqnarray}
\label{diamond}
I_{m,E} & = & I_{m,W} + \frac { I_{m,P} - I_{m,W} } {a}, \nonumber\\
I_{m,N} & = & I_{m,S} + \frac { I_{m,P} - I_{m,S} } {b}, \\
I_{m + \frac{1}{2} ,P} & = & I_{m-\frac{1}{2},P} + \frac { I_{m,P} - I_{m-\frac{1}{2},P} } {c}. \nonumber
\end{eqnarray}
The four principal cardinal directions ($N,E,S,$ and $W$) refer to the cell nodes surrounding $P$ (fig. \ref{cardinal}). 
Indexes $m\pm \frac{1}{2}$ point in the two directions around $m$ in a same latitude ($\xi=C^{te}$).

Direction are swapped with increasing values of $\xi$ and, at a given latitude, with increasing values of $\mu$.
In each direction ($\xi$,$\mu$), a finite volume integration of the BTE yields:

\begin{equation}
\label{ETR}
I_{m,P}= \frac
{ \lambda_W I_{m,W}+ \lambda_{m- \frac{1}{2} } I_{m- \frac {1}{2} ,P} + \lambda_S I_{m,S} + \lambda_o I_{\omega}^o(T_P) }
{ |\mu_m| \Delta_z \frac{r_E}{a} - \frac{\Delta r \Delta z}{w_m} \frac{\alpha_{m+ \frac {1}{2} }}{c} + \lambda_S + \lambda_o },
\end{equation}
with:
\begin{eqnarray}
\lambda_W & = & \Delta z \left(|\mu_m| \frac{r_E}{a} - \mu_m \Delta r \right), \nonumber\\
\lambda_{m-\frac{1}{2}} & = & \frac{\Delta r \Delta z}{w_m c}\left[\mu_m w_m (c-1) - \alpha_{m-\frac{1}{2}} \right],  \nonumber\\
\lambda_S & = & |\xi_m| \Delta r \frac{r_P}{b},  \nonumber\\
\lambda_o & = & r_P \kappa_{\omega} \Delta r \Delta z,
\end{eqnarray}

\begin{equation}
\mbox{and} \quad \nonumber \\
\left\{
\begin{array}{l}
\alpha_{m+\frac{1}{2}}- \alpha_{m-\frac{1}{2}}  = \mu_mw_m, \nonumber\\
\alpha_{\frac{1}{2}}  =  0. \nonumber
\end{array}
\right.
\end{equation}
Interpolation weights $a$, $b$ and $c$ have the following values, which guarantees positivity of the solution while keeping, 
as much as possible, a $2^{nd}$ order accurancy:
\begin{eqnarray}
a &=& \mbox{max} \left (0.5 , 1 - \frac {|\mu_m| r_W \Delta z } 
{\lambda_o + 2|\xi_m| r_P \Delta r  - 2\Delta z \Delta r \frac{\alpha_{m+ \frac{1}{2}}}{w_m}} \right), \nonumber \\
b &=& \mbox{max} \left (0.5 , 1 - \frac {|\xi_m| r_P \Delta r } 
{\lambda_o + 2|\mu_m| r_E \Delta z  - 2\Delta z \Delta r \frac{\alpha_{m+ \frac{1}{2}}}{w_m} }\right), \nonumber \\
c &=& \mbox{max} \left (0.5 , 1 + \frac {\Delta z \Delta r \frac{\alpha_{m-\frac{1}{2}}}{w_m}} 
{\lambda_o + 2|\xi_m| r_P \Delta r  + 2|\mu_m| r_E \Delta z  } \right). \nonumber \\
\end{eqnarray}

As explained by Lemonnier \cite{Lemonnier07}, at each new latitude, for a first direction $m$, the value of $I_{m-\frac{1}{2},P}$, 
corresponding to $\eta=0$, has to be initialised. 
This particular direction is in $(r,z)$ plane. 
Consequently, the angular redistribution contribution ($-\frac{1}{r} \frac{\partial {(\eta I_{\omega,p} )} }{\partial {\phi} }$) is null. 
Therefore, by setting $\alpha_{m-\frac{1}{2}}=0$ and $\mu_m=-\sqrt{1-\xi_m^2}$, $I_{m-\frac{1}{2},P}$ is computed with a and b becoming:
\begin{eqnarray}
a' &=& \mbox{max} \left (0.5 , 1 - \frac {r_W \Delta z \sqrt{1-\xi_m^2} } 
{\lambda_o + 2|\xi_m| r_P \Delta r  +\Delta z \Delta r \sqrt{1-\xi_m^2} } \right), \nonumber \\
b' &=& \mbox{max} \left (0.5 , 1 - \frac {|\xi_m| r_P \Delta r } 
{\lambda_o + (r_E+r_W) \Delta z \sqrt{1-\xi^2} }\right). \nonumber \\ 
\end{eqnarray}

\subsection{Energy conservation}

At this step, we can determine spectral intensities at any nanostructure points, for a given direction $m$, at a frequency $\omega$ and polarisation $p$. 
The integration over the solid angle $d\Omega$ and over the spectrum ($d\omega$ and $p$) yields a total intensity in a point $P$.
To be able to obtain $I_{m,P}$, temperature $T$ is required and is deduced from the energy conservation(Eqn.\ref{NRJ}).

\begin{equation}
\label{NRJ}
\frac{\partial e}{\partial t}+ \nabla\cdot\mathbf{q} =0,
\end{equation}
where $e$ stands for the integrated phonon energy density, and $\mathbf{q}$, the phonon total flux density:
\begin{equation}
	\mathbf{q}(\mathbf{r})=\sum_p\int_0^{\omega_{max}} \int_{4 \pi}I_{\omega,P}(\mathbf{r},\mathbf{\Omega})\cdot\mathbf{\Omega} d\Omega d\omega.
\end{equation}
In steady state, the divergence of the phonon total flux density must be null.
Therefore, integrating Eqn.\ref{Boltzmann} over all directions and the entire spectrum yields equilibrium:
\begin{equation}\label{conservation}
\sum_p \int_0^{\omega_{max}} \int_{4 \pi} \kappa_{\omega} I_{\omega,P} d\Omega d\omega=
\sum_p \int_0^{\omega_{max}} \int_{4 \pi} \kappa_{\omega} I^o_{\omega}(T) d\Omega d\omega.
\end{equation}
This relation is then inverted to get point $P$ temperature from local intesity ($I_{\omega,P}$) knowledge in all direction, 
all frequencies and all relevant branches (Eqn.\ref{conservation}).

\subsection{Dispersion relations}

\begin {figure}[h]
\begin{center}
\includegraphics[width=8.6cm]{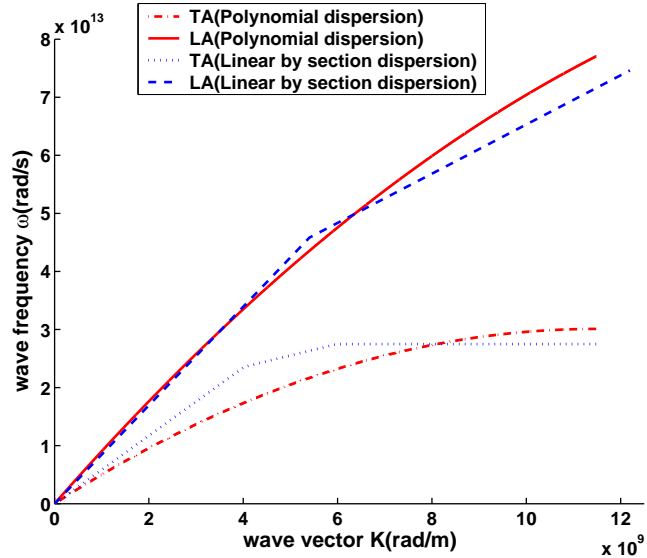}
\end {center}
\caption{(Color online) Silicon dispersion relation in direction (100) given by E.Pop \cite{pop04} }
\label{Si}
\end{figure}

\begin {figure}[h]
\begin{center}
\includegraphics[width=8.6cm]{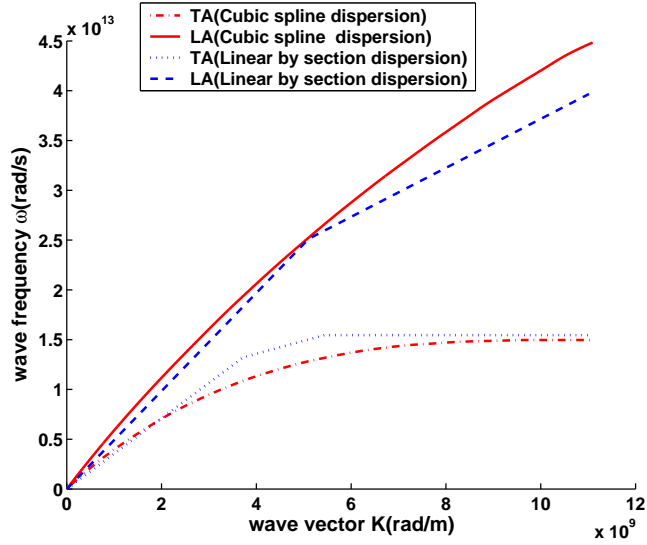}
\end {center}
\caption{(Color online)  Germanium dispersion relation in direction (100) given by G.Nilsson \cite{Nilsson71} }
\label{Ge}
\end{figure}

Many studies have been done for calculating semiconductors thermal properties, using linear \cite{Asen97,Yang05nl,Yang05jht} 
or linear by section\cite{Holland63,Singh03,Liu06} spectral dispersion. 
Fewer litteratures are found with truly non-linear phonon dispersion \cite{Lacroix05}. 
In this work, we use a polynomial phonon dispersion \cite{pop04}, for silicon (Fig.\ref{Si}). 
For each polarisation branch, the wave vector is discretized into $60$ equally spaced values. 
For germanium, as suggested by Lacroix \cite{Lacroix05}, cubic splines are fitted on experimental data \cite{Nilsson71} (Fig.\ref{Ge}).
Germanium wave vector is equally shared into $300$ bands. 
Some assumptions are introduced, semiconductors are assumed as isotropic, where the studied propagation direction is along [100] lattice direction, 
and only bulk acoustical modes are used. 
Optical mode are neglected (their respective branches are ignored).

Real phonon dispersion curves for semiconductor are non-linear. 
Therefore, the group velocity $V_g(\omega,p)=\frac{\partial \omega}{\partial K}$ is dissociated from the phase velocity $V_p(\omega,p)=\frac{\omega}{K}$, 
both being dependent on wave frequency. 
As a consequence, the blackbody intensity (Eqn.\ref{black}) has a complex spectral behaviour. 
The analytical integration over the wave frequency of the equation (\ref{conservation}) is not available. 
To overpass this complex behavior, $V_p(\omega,p)$ is oftenly approximated, in litterature, 
to a constant $V_g(\omega,p)$ (linear by section dispersion curves) or to a constant averaged velocity, whereas, in this work, 
we compute the equilibrium equation. 
As for the phase velocity, 
the group velocity spectral variation is introduced in the extinction coefficient $\kappa_{\omega,p} = \frac {1}{V_g \tau_{\omega,p}}$. 
This last definition shows that the extinction coefficient $\kappa_{\omega,p}$ is a spectral and branch dependent quantity through $V_g(p,\omega)$ and 
the relaxation time $\tau_{\omega,p}$.

Following Matthiessen's rule, $\tau_{\omega,p}$ may be expressed as:
\begin{equation}\label{Matthiessen}
\tau_{\omega,p}^{-1}=\tau_{U}^{-1}(\omega,p)+\tau_{N}^{-1}(\omega,p)+\tau_{i}^{-1}(\omega,p).
\end{equation}
In litterature, Holland relaxation times\cite{Holland63} are oftenly applyed. 
Therefore, as a starting point, we use, in this work, the same form (Tab. \ref{time}) to define impurities, Normal and Umklapp processes\cite{Kittel04}. 
Boundary relaxation time is not used since boundary conditions have been imposed further up. 
Note that the relaxation times can also be temperature dependent.

\section{New phonon relaxation times in dispersive semiconductors}

In this work, the solution of the spectral BTE requires group and phase velocitie determination, deduced from the chosen dispersion curve fit, 
and of the relaxation time approximation. 
This latter is done with Holland formulas. 
A compatibility problem thus occurs since Holland's dispersion curve fits and ours are not similar. 
Therefore, one can wonder about the relaxation time constants $A$, $B_L$, $B_T$, and $B_{TU}$ to input in the numerical simulation. 
For example, the transition between $N$ and $U$ process for transverse phonon is not the same. 
Holland assumes that the cutoff frequency is $\omega_1=\frac{k_B \Theta_1}{\hbar}$ 
whereas Han and Klemens \cite{Han93} state it occurs at $\frac{K_{max}}{2}$.

Thus, these differences bring a new frequency variation and also a new spectral domain. 
Consequently, setting this phonon dispersion to obtain thermal properties with Holland's method yields to an improper answer\cite{Chung04} (Fig.\ref{gros_film} 
and Fig.\ref{gros_film_germanium}). 
Therefore, 
the relaxation times have to be fitted to approach semiconductor conductivity with Holland's relaxation time forms but using non-linear dispersion curves.

\subsection{Silicon}
\subsubsection{Silicon new relaxation times}

Since, 
factors $F$ (boundary) and $A$ (impurity) are present in all the different conductivities $k_{T}$, $k_{TU}$, and $k_L$ defined by Holland \cite{Holland63}, 
we adjust them in a first place using Holland's method to predict conductivity ($k_{T}$ is the conductivity due to impurities, 
boundaries and Normal transverse collisions, $k_{TU}$ is the conductivity due to impurities, boundaries, and Umklapp transverse collisions, 
and $k_L$ accounts with impurities, boundaries, and longitudinal phonons collisions). 
Boundary and impurity interactions are dominant at low temperature \cite{Klemens58,Holland63,Srivastava90,Han93}, 
therefore $F$ and $A$ are fitted in this temperature range.

Around $10$K, close to transverse phonon Normal process highest value, factor $B_{T}$ is fixed. 
At high temperature, the coefficient $B_{TU}$ is set where Umklapp processes for transverse phonons prevails. 
It is unimportant to determine $B_{T}$ before $B_{TU}$, or conversely when one process is dominant the other is negligible. 
On the other hand, factor $B_L$, corresponding to longitudinal contribution is fitted at last, 
hence $k_L$ major portion has transversal conductivities non negligible. 
This last coefficient is found around $100$K.

At very high temperatures, optical phonons are not taken into account. 
Although conductivity values correspond to experimental data, the model does not describe entirely the submitted physics at these temperatures.

\subsubsection{Silicon thick films and silicon thick cylinder}\label{sec_fit}

\begin {figure}[h]
\begin{center}
\includegraphics[width=8.6cm]{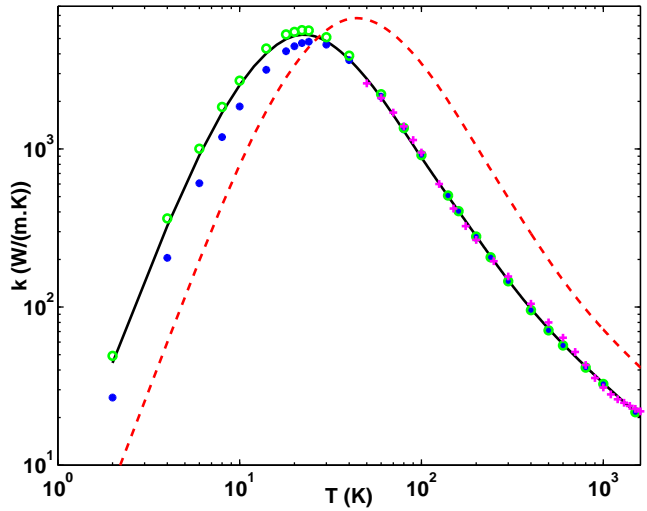}
\end {center}
\caption{(Color online)  This study is done in a silicon structure, where the dimension parameters equal to $7.16$mm.
	The solid line is the conductivity obtain with Holland's method, relaxation times and phonon dispersion. 
	The dot line is the conductivity obtain with Holland's method and relaxation times but with Pop's phonon dispersion.
	Dots ($\bullet$) and ($\circ$) are the conductivity obtain with BTE resolution but with our new relaxation times and Pop's phonon dispersion.
	The wire is corresponding to $\rho=1$  represented by ($\bullet$) and the film to $\rho=0$ drawn with ($\circ$) (Eqn.\ref{reflection}).
    Dots ($+$) are Glassbrenner experimental conductivities.}
\label{gros_film}
\end{figure}

To simulate a bulk using our new parameters (Tab.\ref{tableau}), we settle the cylinder with the film condition ($\rho=0$) for a thickness of $L=7.16$mm. 
In figure (\ref{gros_film}), for temperatures above $100$K, one can notice that the results obtained for the film correspond to Holland's predictions, 
based on measurements\cite{Holland63,Glassbrenner64}. 
For lower temperatures ($T<100$K), the values obtained with BTE resolution are above the reference curve. 
Even if the dimensions are huge, Holland is working on a bulk with an equivalent sample size $L=7.16$mm, 
which appears in the boundary reflection contribution ($\tau^{-1}_b=V_s/LF$). 
In our geometry, the sample size $L$ corresponds more to our diameter $D$. 
Therefore, to simulate a wire, the ratio of diffuse to specular reflection $\rho$ is set equal to one.
Thus, having $D=L=7.16$mm, conductivities obtained for $T>100$K, 
as for the film, match with measurements\cite{Holland63,Glassbrenner64} (Fig. \ref{gros_film}). 
For $T<100$K, the results are now below the curve.

A first set of calculations has been performed with a diffuse to specular reflection ratio $\rho=0$ (film) and has brought conductivities above or equal to the reference curve.
A second set has been examined with $\rho=1$ (wire/cylinder, $D \simeq L$) and has given results below or identical to the curve. 
It can be then assumed that there exists a ratio of diffuse to specular reflection between $0$ and $1$ which will give conductivities similar to experimental data (the geometric parameter $F$ is not use in BTE resolution since the boundary conditions are already made in section \ref{sec_geometry}).

\subsection{Germanium new relaxation times}

\begin {figure}[h]
\begin{center}
\includegraphics[width=8.6cm]{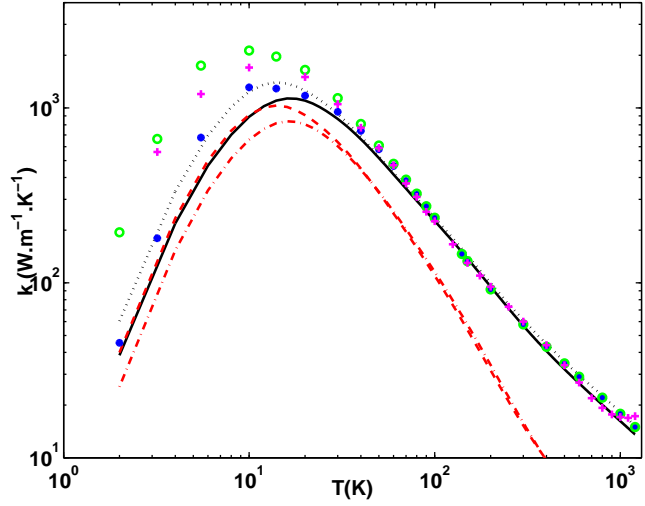}
\end {center}
\caption{(Color online)  This study is done in a Germanium structure.
	The solid line is the conductivity obtained with Holland's method (relaxation times and phonon dispersion), with a sample length of $L=2.4$mm. 
	The dot line is the conductivity obtained with Holland's method but with Asen relaxation time parameters and its dispersion.
	Asen sample length is $L=3.8$mm.
	The dash dot line is the conductivity resulting from Holland's method with Nilsson's phonons dispersion ($L=2.4$mm).
	The hyphen line is the conductivity calculated with Asen parameters on Holland relaxation time forms with Nilsson's dispersion ($L=3.8$mm).
	Dots ($\bullet$) and ($\circ$) are conductivities obtained with BTE resolution but with our new relaxation times and Nilsson's phonon dispersion,
	on a sample of $L=20$mm and $D=4.4$mm. 
      The wire is corresponding to $\rho=1$ represented with ($\bullet$) and the film to $\rho=0$ drawn with ($\circ$) (Eqn.\ref{reflection}).
	Cross dots ($+$) are Glassbrenner experimental data on a sample of $L=20$mm and $D=4.4$mm.}
\label{gros_film_germanium}
\end{figure}

A difficulty is found when we try to fit germanium relaxation time parameters with the equations employed by Holland.
The obtained results always underestimate conductivity (Fig.\ref{gros_film_germanium}), 
with $L=D=2.4$mm (Holland's equivalent sample length and relaxation parameters) or with $L=D=3.8$mm 
(Asen-Palmer's equivalent sample length and relaxation parameters).
Germanium thermal property alters with various models and experimental data.
Varying the sample length modifies conductivity values at low temperature where boundary scattering is important. 
Furthermore, germanium conductivity seems to be sensible to the material doping \cite{Geballe58,Bird71,Asen97,Singh03} which increases considerably conductivity curve highest point. 
Thus, changing the impurety parameter $A$ alters considerably the conductivity around $10$K. 
Fitting then the three first parameters $F$, $A$, and $B_T$ to obtain correct values for low temperature undergo a mismatch around $140$K to $300$K.

To avoid this trouble, Singh \cite{Singh03} proposes different relaxation time forms based on  three different sections of the dispersion curves.
Here, as for silicon, we want to use Holland's relaxation time forms (Tab.\ref{time}) to model germanium thermal properties, 
where only four parameters are needed against $10$ for Singh. Therefore, we solve directly the BTE to fit our parameters. 
Inquired values are based on Glassbrenner and Slack experimental data\cite{Slack60, Holland63, Glassbrenner64}. 
The cylinder dimensions have the same lengths than their sample ($L=20$mm and $D=4.4$mm).

A first fit is done for $T \geq 100$ K, where conductivity is less sensible to boundary collisions and to Normal transversal processes.
Thus, $B_{TU}$, $B_L$ and $A$ parameters are the conductivity action switchs over $100$K. 
The dominant band for tranverse normal process conductivity contribution is under $100$K, 
which permits to change $B_T$ without disturbing our new values obtained for $T \geq 100$K.

These new germanium relaxation time parameters are given in Table \ref{tableau_Ge}. 
The resulting conductivities have less than $5\%$ of relative difference with experimental data for $T\geq 100$K (Fig. \ref{gros_film_germanium}). 
Just below $100$K, a maximum of $7\%$ of relative difference is seen. 
Under $60$K, Glassbrenner's conductivities are always between our film and wire values. 
One can say that there exists a single ratio of diffuse to specular reflection, for each temperature, 
which will give conductivities similar to these experimental data. Therefore we are able to valid our germanium parameters.

Note that the spectral discretization is finer for germanium than for silicon. 
The first Brillouin zones are of the same order, but silicon wave vector is divided into $60$ equal bands whereas, for germanium, it is splitted into $300$ strips. 
Using $300$ bands for silicon changes conductivity values by less than $1\%$. 
Therefore calculating over $60$ wave frequencies saves calculation times while keeping accuracy to an acceptable level. 
On the other hand, using only $60$ bands over the spectrum can change by up to $10\%$ of the conductivity values. 
To take a maximum of spectral information, we have choosen a large number ($300$) of bands.

\section{Heat transfer in silicon nanostructures}

\subsection{Films}
New parameters (Tab.\ref{tableau}) have been fitted to provide correct thermal properties in silicon 
and they are now used to study films with BTE resolution. 
As said earlier, BTE gives the possibility to study ballistic, mesoscopic and diffusive problems. 
Therefore, a thickness variation brings us to study nanofilms to thick films (or bulk). 
Ballistic phenomena is then more expected for nanofilms whereas diffusion occurs in thick films. 
Another way to observe ballistic behavior is to work with low temperatures. 
In these condition, low frequency modes are dominant (due to Bose-Einstein distribution) 
which means that phonon mean free path is larger (due to the relaxation time frequency dependence). 
On the other hand, at high temperatures, high frequency modes are dominant, which amounts to small phonon mean free path. 
Three phonon collisions are then more frequent, which drive the process to a diffusive scheme.

\begin {figure}[h]
\begin{center}
\includegraphics[width=8.6cm]{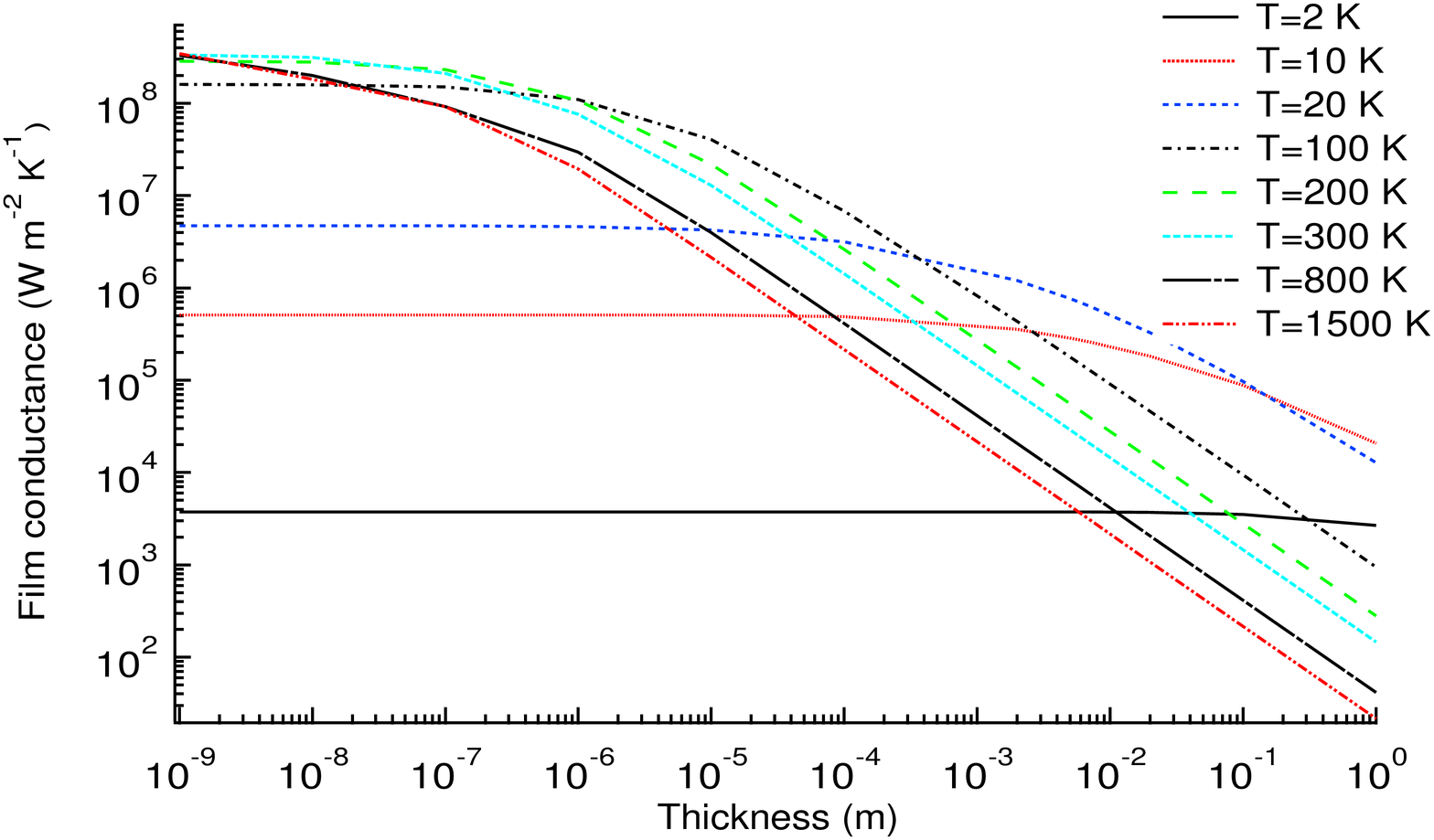}
\end {center}
\caption{(Color online)  Film surface unit conductance versus thickness at different temperature.}
\label{film}
\end{figure}

Figure \ref{film} shows silicon surface unit conductance (W/m$^2$K) versus its film thickness for several temperatures.
According to the film assumption, the section perpendicular to $\mathbf{z}$ is supposed to be infinite. 
Therefore, conductance (W/K) is not available. In a film, the surface unit conductance $G"$ reads:
\begin{equation}
\label{cond}
G"  = \frac{\varphi}{-\Delta T}=  \frac{k}{\Delta z}.
\end{equation}

Furthermore, in diffusive regime, when Fourier's law (Eqn.\ref{Fourier}) is matching the heat transfer, the conductivity, at a given temperature, is constant. 
In that case we get $\ln(G") =  -\ln(\Delta z)+\ln(k)$.
This result correspond to the right part of Figure(\ref{film}).

For small lengths and at a given temperature, the surface unit conductance becomes constant when thickness becomes thinner. 
This scheme is due to ballistic phenomena and is viewed on curves left side. 
On the contrary to Fourier's regime, when $G"$ is constant, 
the conductivity concept is questionnable since it depends on the system size (Eqn.\ref{cond}).

The curved parts of $G"$ (Fig.\ref{film}) are the results obtained in mesoscopic regimes. 
In this scheme, phonon mean free path is of the same order than the film thickness. 
Some phonons act as they were in a diffusive system. 
On the other hand, many phonons have a purely ballistic behavior.
Consequently, the thermal conductivity concept is again questionnable.

\begin {figure}[h]
\begin{center}
\includegraphics[width=8.6cm]{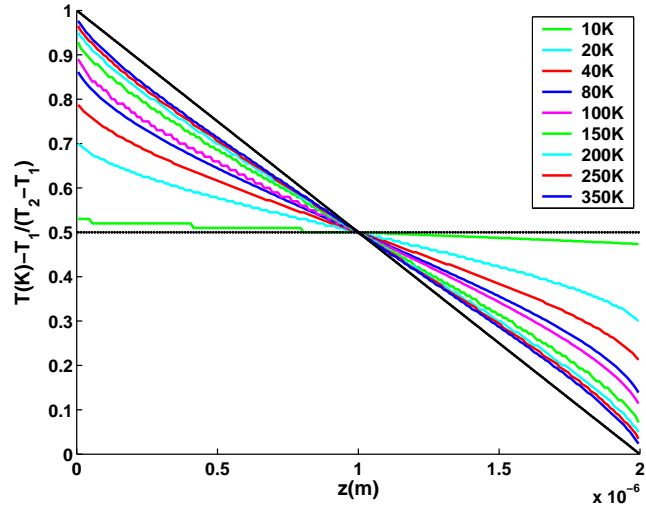}
\end {center}
\caption{(Color online)  Axis temperature field in a $2$micron silicon film.}
\label{champ_film}
\end{figure}

Another benefit in using BTE resolution is that temperature fields in nanostructures can be obtained.
In a diffusive regime, the temperature gradient is constant. In ballistic regime, it is the temperature which is mostly constant.
Figure \ref{champ_film} shows thermal profiles along the $\mathbf{z}$ axis for a two-micron-thick-film. 
We can clearly see that Fourier's regime is not reached before $300$K, which correspond almost to the right mesoscopic limitviewed on Figure (\ref{film}). 
Ballistic regime is nearly reached at $10$K. Between these two temperatures, mesoscopic regime prevails. 
This result confirms that thermal conductivity terms have to be used with care. 
It seems that in silicon films, it is not appropriate to use this concept below the micrometer scale.

\begin {figure}[h]
\begin{center}
\includegraphics[width=8.6cm]{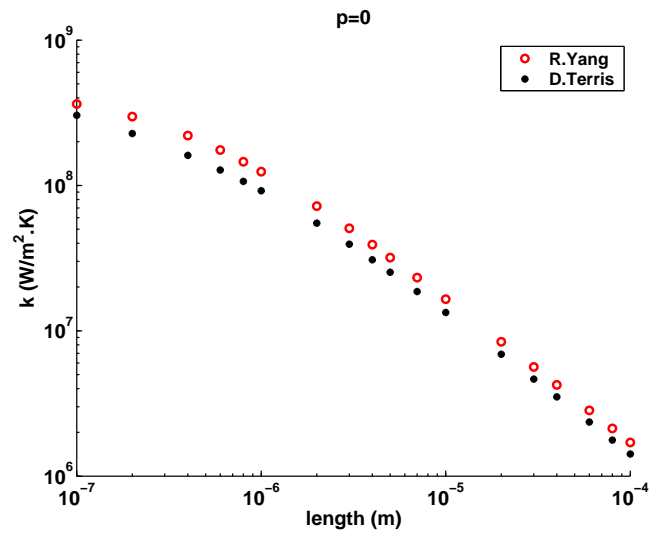}
\end {center}
\caption{(Color online)  Surface unit conductance comparison between a simple model(Yang) and our model(Terris), of films set at 300K.}
\label{Chen_film}
\end{figure}

In comparison with a simple model \cite{Yang05nl}, the surface unit conductance, calculated with a single linear group velocity, 
is overestimated compared to our model (Fig.\ref{Chen_film}). At $300$K, some points are $30\%$ higher than ours for the film simulation. 
These results show how important is to take into account the spectral dependency. 
Consequently phonon mean free paths ($\ell$) are also spectrally dependent since $\ell= V_g(\omega,p) \tau({\omega,p})$.

Note that for a thickness beneath $20.10^{-9}$m, there are less than $40$ primitive cells across the nanofilm.
In that case, bulk dispersion properties are, in principle, no longer valid. 
Correspondant results for these thicknesses are given (Fig.\ref{film}) only to show the numerical behavior of our code despite a physical concordance.
 
\subsection{Wires}

\begin {figure}[h]
\begin{center}
\includegraphics[width=8.6cm]{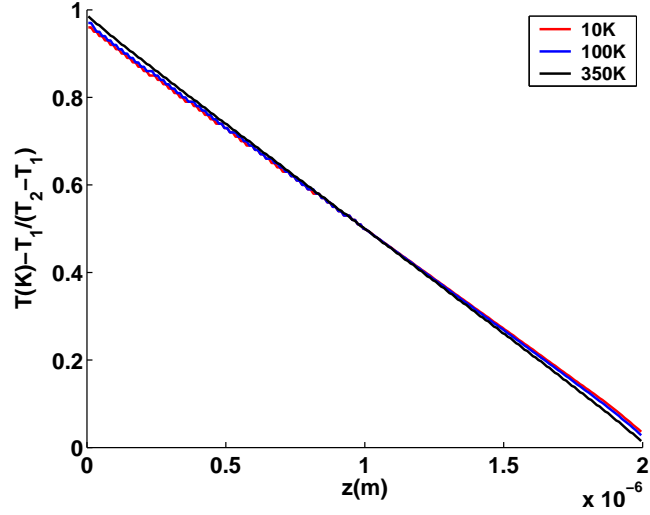}
\end {center}
\caption{(Color online)  Axis temperature field in a $2\mu$m length and $115$nm diameter nanowire.
		The ratio of diffuse to specular reflection is equal to $1$.}
\label{champ_fil}
\end{figure}

Silicon new relaxation time parameters are used to determine nanowire thermal properties. 
In comparison with nanofilms, speaking about thermal conductivity in nanowire could be taken out of sense. 
Nevertheless, representing the surface unit conductance versus the length do not present much interest for wires, 
since the thermal property changes also with its diameter. 
Therefore, we plot the temperature field to see, with the thermal gradient along the $\mathbf{z}$ axis, in which regime, ballistic, 
mesoscopic or diffusive, the nanowire, at a given temperature, is. 
For a $2$micron long nanowire whose  diameter is $115$nm, Fourier's law applies on a large temperature band (Fig.\ref{champ_fil}). 
Using thermal conductivity concept to describes nanowire thermal properties seems then possible.

A temperature rise leads to Umklapp process domination which is a resistive process and then yields a diffusive regime. 
Similarly, a diameter reduction increases phonon collisions with the border, also a resistive process, and favours a diffusive regime.
Therefore, thermal conductivity as silicon nanowire thermal property is suitable for larger temperature or smaller diameter.

For a given diameter, a shorter wire gives more possibility to have phonon mean free path of the same order as the wire length. 
Low temperature can also accentuate this character as phonon traveling length grows up.
Other results have been obtained with $D=500$nm and $L=2\mu$m, or with $D=37$nm and $L=150$nm. 
Their temperature fields have the same shape than those shown in figure (\ref{champ_fil}). 
This tends to consider that thermal conductivity is a valid concept in nanowires. 
However mesoscopic and ballistic regime can always be obtained in any nanostructures by reducing hugely the dimensions or/and the temperature.

\begin {figure}[h]
\begin{center}
\includegraphics[width=8.6cm]{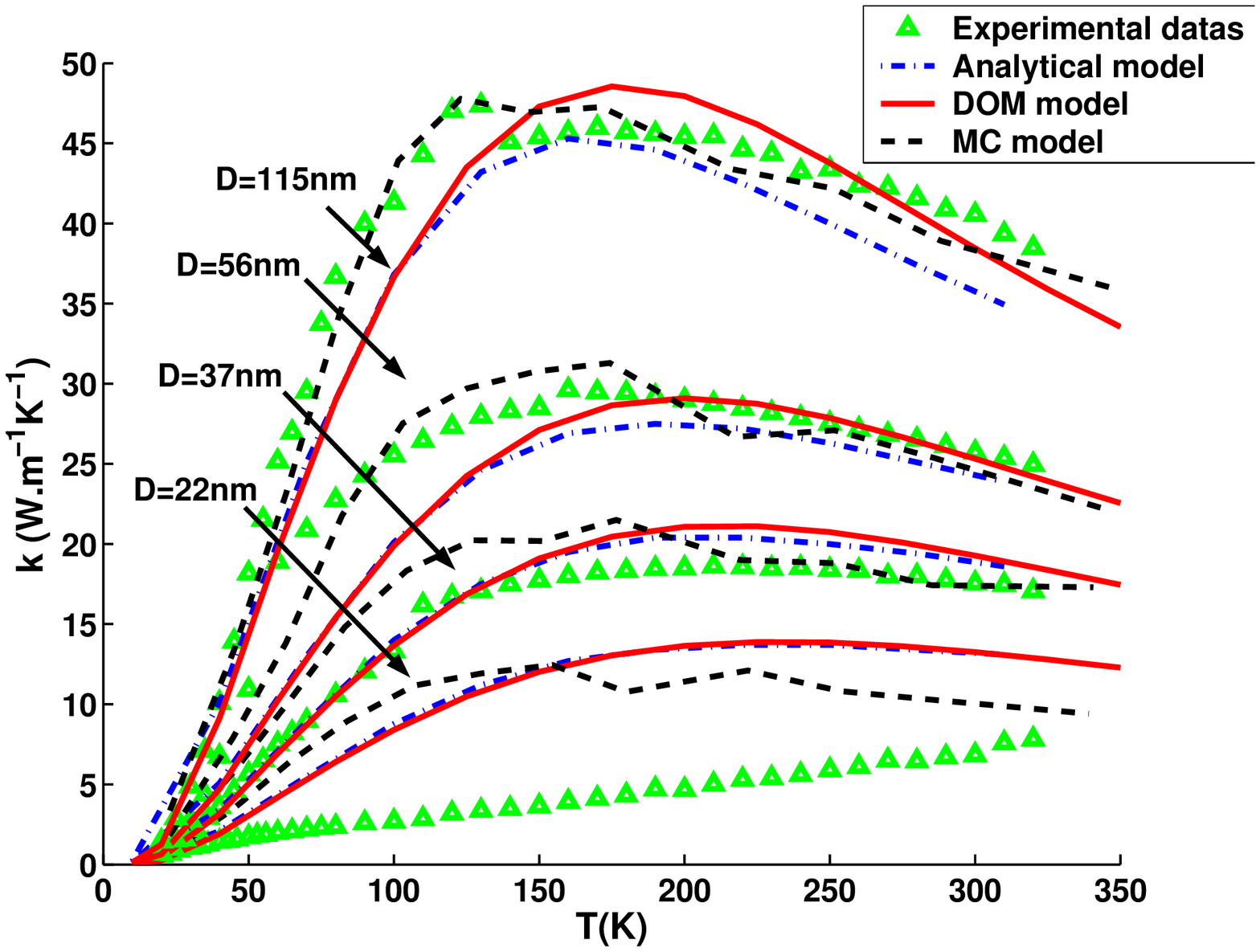}
\end {center}
\caption{(Color online)  Nanowire themal conductivity. 
		Comparison with experimental results of Li et al. \cite{Li03}, 
		analytical data of Chantrenne et al. \cite{Chantrenne05} and Monte Carlo simulations of Lacroix et al. \cite{Lacroix06}}
\label{k_fil_p1}
\end{figure}

We have calculated the conductivity for $2\mu$m length nanowires at diffenrent temperatures and diameters (Fig.\ref{k_fil_p1}). 
We retrieve a same profile than analytical \cite{Chantrenne05}, experimental \cite{Li03} and Monte Carlo \cite{Lacroix06} data. 
However, a better agreement is found at high temperature for large diameters and at low temperature for small diameters. 
The boundary has been set here as purely isotropically scattering surface.

\begin {figure}[h]
\begin{center}
\includegraphics[width=8.6cm]{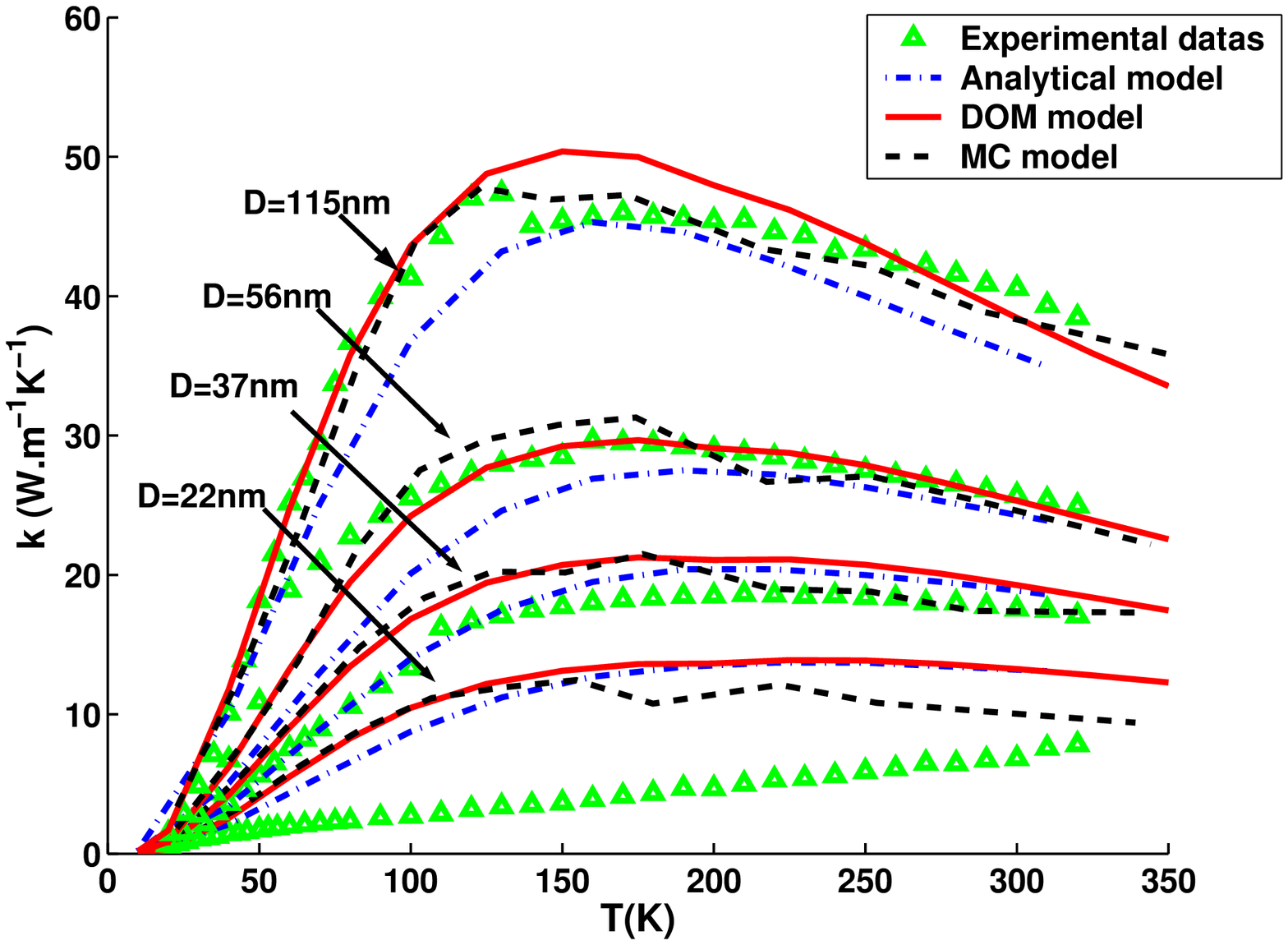}
\end {center}
\caption{(Color online)  Nanowire themal conductivity. 
		Comparison with experimental results of Li et al. \cite{Li03}, 
		analytical datats of Chantrenne et al. \cite{Chantrenne05} and Monte Carlo simulations of Lacroix et al. \cite{Lacroix06}}
\label{k_fil_p08}
\end{figure}

As seen in section \ref{sec_fit}, a diffuse to specular reflection ratio $\rho$  smaller than one is expected for low temperatures. 
A fit on $\rho$ (Tab.\ref{rho}), is then done on our $2\mu$m long and $115$nm of diameter nanowire.
It is used for other diameters (Fig.\ref{k_fil_p08}). 
Conductivity matches better at low temperature for large diameters, with an adjusted $\rho$ than with $\rho$ fixed at $1$. 
Note that the ratio decreases with temperature. 
When temperature goes down, low frequency phonons are dominant, 
and their wavelength average increases (due to Bose-Einstein distribution). 
Consequently, phonons will scatter less with the borders since some of their wavelengths will be larger than the nanowire roughness. 
Therefore, the ratio of specular to diffuse reflection goes down with temperature.

On the other hand, for small diameters, the conductivity obtained does not match well experimental data.
This can be explained by the fact that with a small diameter, bulk dispersion data are no longer pertinent. 
At small scale, phonon modes become discrete whereas bulk dispersion is continue.
The contribution of non-existing modes probably overestimates heat transfer.

For a $37$nm diameter nanowire with a ratio of specular to diffuse reflection set equal to $1$, 
the difference between our model predictions and Yang's simple model \cite{Yang05nl} are below $7\%$, but with $\rho=0.8$, ot reaches $13\%$.
For every thin wire, boundary scattering has an important resistive role which favours diffusive regime. 
If we further change the boundary conditions, a larger difference is obtained between our model and Yang's. 
Therefore, neglecting the spectral dependency can yield a rough answer.

\section{Conclusion}

This paper presents the resolution of Boltzmann Transport Equation with the Discrete Ordinate Method in semiconductors. 
Spectral and temperature dependencies have been taken into account to define thermal properties.
The polynomial function and the cubic spline approximations, representing frequencies $\omega$ versus wave vector, bring us closer to reality.
Different relaxation time models, which are made to take into account spectral dependency, are proposed in litterature.
We have focused on the same relaxation time forms applied by Holland, which is oftenly used in literature.
Thus, new relaxation times have been found to balance non-linear dispersion relations.
Sensibilities over parameters defining semiconductor relaxation times are different for silicon and germanium.
Therefore, for each semiconductor, a strategy has been developped to find its own parameters.
A comparison with a simple model has put in forward the spectral dependency problem. Differences over $30\%$ can be obtained.
It seems that simple models are correctly fitted for specific samples or temperatures.
Our new parameters, associated to this paper model, 
permit to describe correctly germanium and silicon thermal properties over a large temperature band and over a large spatial scale.
However, this method does not treat optical phonons and therefore it does not represent properly high temperature phenomena.
Futhermore, the dispersion relations are taken from the bulk modes. 
Working on very short scales can undergo incorrect answers since phonon dispersion changes.

Silicon nanofilm surface unit conductance as silicon nanofilm field temperature have shown that thermal conductivity is not always a relevant quantity, 
since diffusive regime is only reached, at $1500$K, for one micron thick. 
Even if optical phonons are not described, it had been viewed that at lower temperatures diffusive regime is only reached for thick films.
On the contrary, we have seen that thermal conductivity is a relevant quantity in nanowires due to phonon collision with borders. 
In extremely small nanostructures, we have noticed that our treatment can be inproper. Phonon wave behaviour should probably be taken into account. 
Note that the method developed here can easily be generalized for transient heat transfer in order to study heat pulse 
propagation in nanostructures\cite{Terris07}. 


\begin {table}[h]
\begin{center}
\begin {tabular}{||c||c||}
\hline
Impureties         & $\quad\tau_i^{-1} $\\
(Rayleigh's model) & $ = A  \omega^4$ \\
\hline
Longitudinal       & $\quad\tau_{Normal+Umklapp}^{-1}$\\
polarization       & $ =  B_L \omega^2 T^3\qquad$\\
\hline
                   & $\quad\tau_{Normal}^{-1}$\\
Transversal        & $=B_T \omega T^4 \;{\rm if }\; 0\leq K <  \frac{K_{max}}{2}$\\
\cline{2-2}
 polarisation      & $\quad\tau_{Umklapp}^{-1}$\\
	             & $=B_U \frac{\omega^2}{\sinh \left(\frac{\hbar \omega}{k_BT}\right)}\;{\rm if}\;\frac{K_{max}}{2}\leq K \leq K_{max}$\\
\hline 
\end{tabular}
\end {center}
\caption{Relaxation time forms.
       ($K$ is the wave vector, with $K_{max}$ corresponding to the first Brillouin's zone, and $k_B$ is Boltzmann's constant.
       $K_{max}=\frac{2 \pi}{a_o}$, where the pure lattice parameter $a_o=543$pm for silicon\cite{Virginia,MacKee01} 
       and $a_o=565$pm for germanium\cite{MacKee01}.)}
\label{time}
\end{table}

\begin {table}[h]
\begin{center}
\begin {tabular}{||p{2cm}||*{2}{c|}|}
\hline
                    & Holland           & Present work  \\
\hline
\hline
 $F$                & 0.8               & 0.66 \\
\hline
 $A$ (s$^3$)        & 1.32 10$^{-45}$   & 1.498 10$^{-45}$ \\
\hline
 $B_{T}$ (K$^{-3}$) & 9.3 10$^{-13}$    & 8.708 10$^{-13}$ \\
\hline
 $B_{TU}$ (s)       & 5.5 10$^{-18}$    & 2.89 10$^{-18}$ \\
\hline
 $B_L$ (K$^{-3}$)   & 2.0 10$^{-24}$    & 1.18 10$^{-24}$ \\
\hline
\end{tabular}
\end {center}
\caption{Parameters for silicon relaxation times}
\label{tableau}
\end{table}

\begin {table}[h]
\begin{center}
\begin {tabular}{||p{2cm}||*{3}{c|}|}
\hline
				& Holland 		& Asen-Palmer		& Present work  \\
\hline
 $L$ (mm)			& $2.4$		& $3.8$			& $D=4.4$ and $L=20$ \\
\hline
\hline
 $F$ 				& $0.8$ 		& $0.8$			& none \\
\hline
 $A$ (s$^3$)		& $2.4 10^{-44}$  & $1.7786 10^{-44}$	& $3.5 10^{-45}$ \\
\hline
 $B_{TO}$ (K$^{-3}$)	& $1.0 10^{-11}$ 	& $1.5 10^{-11}$		& $7.3 10^{-11}$ \\
\hline
 $B_{TU}$ (s) 		& $5.0 10^{-18}$ 	& $4.5 10^{-18}$		& $0.89 10^{-18}$ \\
\hline
 $B_L$ (K$^{-3}$)		& $6.9 10^{-24}$ 	& $9.0 10^{-24}$		& $8.6 10^{-24}$ \\
\hline
\end{tabular}
\end {center}
\caption{Parameters for germanium relaxation times}
\label{tableau_Ge}
\end{table}

\begin {table}[h]
\begin{center}
\begin {tabular}{||p{1cm}||*{12}{c|}|}
\hline

$\rho$	& 0.8 & 0.81 & 0.82 & 0.83 & 0.8 & 0.86 & 0.9 & 0.94 & 0.97 &   1  \\
\hline
$T$ (K)	& 10  & 20   & 40   &  60  & 80  & 100  & 125 & 150  & 175  & $\geq$200  \\
\hline
\end{tabular}
\end {center}
\caption{Ratio of specular to diffuse reflection used in figure \ref{k_fil_p08} for silicon nanowires}
\label{rho}
\end{table}

\end{document}